\documentclass[twocolumn,aps,prd,amsmath,amssymb]{revtex4}
\usepackage{epsfig,bm,dcolumn,color}
\usepackage{graphicx}
\newcommand {\beq} {\begin{equation}}
\newcommand {\eeq} {\end{equation}}

\newcommand {\up}  {\ensuremath{\uparrow}}
\newcommand {\dn}  {\ensuremath{\downarrow}}
\newcommand {\bqa} {\begin{eqnarray}}
\newcommand {\eqa} {\end{eqnarray}}

\newcommand {\kk} {\ensuremath{{\bf k}}}

\newcommand {\qq} {\ensuremath{{\bf q}}}

\begin{document}
\title{Theory of Radio Frequency Spectroscopy of Polarized Fermi Gases}
\author{William Schneider$^{(1)}$}
\author{Vijay B.~Shenoy$^{(1,2)}$}
\author{Mohit Randeria$^{(1)}$} 
\affiliation{(1) Department of Physics, The Ohio State University, Columbus, Ohio 43210}
\affiliation{(2) Department of Physics and Center for Condensed Matter Theory, Indian Institute of Science, Bangalore}
\begin{abstract}

We present two exact results for singular features in the radio frequency intensity $I(\omega)$ 
for ultracold Fermi gases. First, in the absence of final state interactions, 
$I(\omega)$ has a universal high frequency tail $ C\omega^{-3/2}$ for \emph{all} many-body states,
where $C$ is Tan's contact.
Second, in a \emph{normal} Fermi liquid at $T=0$, $I(\omega)$ has a jump discontinuity of
$Z/(1 - m/m^{*})$, where $Z$ is the quasiparticle weight and $m^*/m$ the mass renormalization. We then describe various approximations for $I(\omega)$ in polarized normal gases. We show why an approximation
that is exact in the $n_\dn=0$ limit, fails qualitatively for $n_{\dn} > 0$: there is no universal tail 
and sum rules are violated. The simple ladder approximation is qualitatively correct for very
small $n_{\dn}$, but not quantitatively.
\end{abstract}
\maketitle 


There has been intense experimental activity on characterizing various states of 
matter in ultracold atomic gases \cite{bloch_rmp,stringari_rmp}.
This will become ever more important with the possibility of new and exotic states 
being realized in these systems.
An important tool in these studies is radio frequency (RF) spectroscopy where an RF
pulse is used to transfer atoms from one hyperfine level to another. The RF signal
\cite{chin,schunk} 
has turned out to be much harder to interpret than initially thought because of 
complications of strong final state interactions and the inhomogeneity of trapped gases. 
Recently it has become possible to eliminate these problems by choice of suitable hyperfine levels 
(in $^6$Li) and by tomographic techniques that focus on specific regions of the gas. The most detailed
experimental results are available for polarized Fermi gases 
\cite{ketterle2008,zwierlein2009}.

Motivated by these experiments, we first describe two exact
results for the \emph{singular} features in RF spectra of a two-component Fermi gas with arbitrary interactions.
Our results complement the exact results on sum rules \cite{baym,zwerger}.
We work in the limit where final state interaction effects are negligible,
so that we can focus on the nontrivial effects of interactions in the many-body state.
\\
(1) The RF spectrum $I(\omega)$ has a universal $C {\omega}^{-3/2}$ tail at high frequencies,
where $C$ is Tan's contact coefficient \cite{tan}, which is independent of spin. This form is valid for
\emph{all} phases of Fermi gases: superfluid \cite{bloch_rmp,stringari_rmp}, 
highly imbalanced normal Fermi liquid \cite{zwierlein2009,zwierlein2006,hulet2006,giorgini,combescot} or even
a balanced Galitskii Fermi Liquid \cite{agd}. 
\\
(2) In any normal Fermi liquid state, the RF spectrum $I(\omega)$ at $T=0$ has a jump discontinuity.
Its location depends on the chemical potential $\mu$ and its
magnitude is determined by the combination of Fermi liquid parameters $Z/(1 - m/m^*)$,
where $Z$ is the quasiparticle weight and $m^*$ the effective mass.

These exact results are important not only in interpreting experiments, but also 
in understanding various approximation schemes \cite{zwerger,sachdev,perali},
which are necessarily required to calculate the RF lineshape $I(\omega)$ for 
a strongly interacting gas. In the second part of our paper we critically analyze diagrammatic
approximations for the highly imbalanced normal Fermi liquid.
\\
(3) We show that a simple self-consistent approximation, motivated by 
the fact that it is essentially exact in the $n_{\dn} = 0$ limit \cite{combescot},
has serious qualitative problems for non-zero $n_{\dn}$:
The minority spins do not exhibit the universal tail leading to sum rule violations
and majority spins are completely unaffected.
\\
(4) A simple ladder approximation, on the other hand, correctly exhibits all of the qualitative features
expected on general grounds for $n_{\dn}>0$, however there are quantitative inaccuracies and the
approximation breaks down for $n_{\dn} \gtrsim 0.05$. 

{\bf Formalism:} Consider a Fermi gas with three hyperfine states which we label as $\up$, $\dn$ and ``$e$'' (for excited or empty).
The number density in levels $\sigma = \up,\dn$ is $n_\sigma$ with corresponding (non-interacting) Fermi energies 
$\epsilon_{F\sigma}$. The $\up$ and $\dn$ fermions interact with an s-wave scattering length $a_{\up,\dn} \equiv a$.
The $e$-level is located at energy $\Delta E_\sigma$ from the bottom of the $\sigma$ bands,
and is empty ($n_e = 0$). We assume that fermions in the $e$ 
state do not interact with those in $\sigma$ levels: $a_{e,\sigma} \equiv 0$. If such interactions are strong,
the simple results obtained below are considerably modified by vertex corrections \cite{perali}. One is then 
dealing with the complications of the probe in addition to the many-body system of interest.

When final state interactions are negligible
linear response theory leads to the simple result:
\beq
I_\sigma(\omega) = \sum_{\kk}A_\sigma(\kk,\epsilon_\kk - \mu_\sigma - \omega)n_F(\epsilon_\kk - \mu_\sigma - \omega)
\label{Iw}
\eeq
where the RF~shift $\omega = \omega_{RF} - \Delta E_\sigma$, $\omega_{RF}$ is the RF frequency,
$\epsilon_\kk = k^2/2m$ is the bare dispersion, $\mu_\sigma$ the chemical potential, and $\hbar = 1$.
$n_F(\epsilon)$ is the Fermi function and the single particle
spectral function $A_\sigma(\kk,\omega) = - {\rm Im}G_\sigma(\kk,\omega+i0^+)/\pi$ includes all many-body
renormalizations due to interactions between $\up$ and $\dn$ fermions. 

{\bf Sum Rules and Large-$\omega$ behavior:} 
The exact sum rules \cite{baym,zwerger} for the zeroth $(\ell = 0)$ and 
first $(\ell = 1)$ moments of the RF intensity $\int d\omega \omega^\ell I_\sigma(\omega)$,
are valid for all values of $a$ and $a_{e,\sigma}$. 
It might seem that the first moment sum rule (clock shift),
which diverges as $a_{e,\sigma} \to 0$, can be of no
use when final state interactions are negligible.  However, we find that this divergence is
actually related to a universal high frequency tail in $I_\sigma(\omega)$. 

We rewrite (1) as
$I_\sigma(\omega) = \sum_{\kk} \int d\Omega A_\sigma(\kk,\Omega)n_F(\Omega)
\delta(\Omega - \epsilon_\kk + \mu_\sigma + \omega)$.
This immediately leads to zeroth moment sum rule \cite{normalization}
$\int d\omega I_\sigma(\omega) = N_\sigma$,
using $\int d\Omega A_\sigma(\kk,\Omega)n_F(\Omega) = n_\sigma(\kk)$ and $\sum_\kk n_\sigma(\kk) = N_\sigma$ is the number of $\sigma$ fermions.

We next analyze $I_\sigma(\omega \to \infty)$.  
The delta-function $\delta(\Omega - \epsilon_\kk + \mu_\sigma + \omega)$ then
contributes in one of two ways: either (a) $\Omega$ is large negative with $\epsilon_\kk$ small, or 
(b) $\Omega$ small but $\epsilon_\kk$ large. In case (a), however,
the spectral function $A_\sigma$ vanishes for small $\kk$ and $\Omega \to -\infty$. 
Thus only case (b) survives and we find
$I_\sigma(\omega \to \infty) \simeq \sum_{\kk} n_\sigma(\kk)\delta(\epsilon_\kk - \omega)$.
Using Tan's result \cite{tan} $n_\sigma(\kk) \simeq C/k^4$ for $k \gg k_F$ we thus find that
\beq
I_\sigma(\omega \to \infty) \approx \frac{1}{4 \pi^2 \sqrt{2m}} C \omega^{-3/2},
\eeq
where $C$ is the contact.
We emphasize that the form of this result is independent of the phase (normal or superfluid) of the Fermi gas,
though the value of $C$ does depend on the phase. (This tail is absent only for the \emph{noninteracting} gas
for which $C \equiv 0$.) Note that this high frequency tail arises from
short-distance physics in any Fermi gas, and is
crucial for enforcing the \emph{divergent} clock shift for $a_{e,\sigma} = 0$.

{\bf Fermi liquid singularity:} In the study of many-body systems, various phases are often
directly identified by characteristic low-energy singularities in measurable quantities, such as the 
the discontinuity at $k_F$ in the momentum distribution of a Fermi liquid, or the square root 
singularity in the density of states of a s-wave superconductor at $T=0$.
Here we ask if any such singularity exists in the RF signal. Given the $\kk$-sum and the
kinematics in eq.~(1), we see that there is no characteristic singularity in 
the paired superfluid state. However, as we show next, there \emph{is} a singular signature
for normal Fermi liquids at $T=0$.

In the remainder of this paper we focus on the \emph{normal} (i.e., non-superfluid)
ground state of the highly polarized Fermi gas. Thus our results are relevant, e.g., to 
the unitary gas which has been predicted to be a normal Fermi liquid 
for $x = n_{\dn} / n_{\up} < 0.4$, based on quantum Monte Carlo simulations \cite{giorgini}. 
(Our general results apply equally well to the dilute repulsive gas of Galitskii \cite{agd},
which is yet to be realized in the laboratory.)

For a Landau Fermi liquid the spectral function is
\beq
A(\kk,\omega) \simeq Z \delta\left(\omega - k_F(k - k_F)/m^*\right) + A^{\rm inc}(\kk,\omega)
\label{Afl}
\eeq
close to the Fermi surface $(k\simeq k_F,\omega \simeq 0)$.
The subscript $\sigma$ is dropped for simplicity. The first ``coherent'' term gives
the quasiparticle pole in the Green's function with quasiparticle weight $Z$ and effective mass $m^*$ \cite{FLresults}.
$k_F$ is unshifted from its bare value as required by Luttinger's theorem \cite{agd}.
The second non-singular term is the ``incoherent'' part of the spectral function.

The \emph{singular} contribution to $I(\omega)$ is obtained by substituting the coherent term in 
(\ref{Afl}) into (\ref{Iw}) and using $n_F(\epsilon) = \Theta(-\epsilon)$ at $T=0$.
We convert the $\kk$-sum to an integral over $\epsilon_\kk$ and write the quasiparticle dispersion
as $k_F(k - k_F)/m^* \simeq (k^2 - k_F^2)2m^* = (\epsilon_\kk - \epsilon_F)m/m^*$.
For $m^* > m$ we find a peak which grows like a square root in $\omega$ and then 
has discontinuous drop, all of which rides on top of 
top of the smooth contribution from the incoherent piece.
The location of the discontinuity $\omega^*$ and the size of the jump $\Delta I$ are thus given by \cite{note}
\beq
\omega^*_{\sigma} = \epsilon_{F\sigma} - \mu_{\sigma}; \ \ \ \ \ 
\Delta I_\sigma = {Z_\sigma N(\epsilon_{F\sigma}) \over \left(1 - m/m^*_\sigma\right)},
\label{disc}
\eeq 
where $N(\epsilon_{F\sigma})$ is the density of states at the Fermi energy.

{\bf Diagrammatic Lineshape Calculations:} The form of the Fermi surface singularity and the high energy tail 
in the RF~intensity have been elucidated above on general grounds. 
Calculating the detailed lineshape $I(\omega)$ necessarily requires approximations to be made
for a strongly interacting Fermi system.
Here we describe diagrammatic calculations for the highly imbalanced normal gas,
highlighting the successes and limitations of two approximation schemes. All such calculations sum
particle-particle (p-p) channel ladder diagrams:
$\Gamma^{-1}(\qq,iq_\ell) = m/4\pi a - \sum_\kk \left[1/2\epsilon_\kk - 
  \beta^{-1}\sum_{n} G_\up(\kk+\qq,ik_n + iq_\ell)G_\dn(-\kk,-ik_n)\right]$.
The rationale for focusing on p-p ladders can be given in many different ways. These are the 
leading diagrams for Fermi systems with short-range interactions 
in the low density repulsive Fermi liquid \cite{agd}, in the
Nozieres-Schmitt-Rink analysis \cite{nsr} of the normal state of the BCS-BEC crossover, 
and in the $1/N$ expansion for the attractive Fermi gas \cite{sachdev}. 

One can analytically obtain closed-form
expressions for the real and imaginary parts of the retarded $\Gamma^{-1}(\qq,\omega+i0^+)$
when $G_\sigma$ are the \emph{bare} Green's functions; details are omitted for
simplicity \cite{nikolic}. Next, the self-energies 
$\Sigma_\sigma(\kk,ik_n) = \beta^{-1}\sum_{\qq,\ell} \Gamma(\qq,iq_\ell)G_{-\sigma}(-\kk+\qq,-ik_n + iq_\ell)$
are calculated \cite{details} using the spectral representation for the p-p vertex in terms of 
${\rm Im}\Gamma(\qq,\omega+i0^+)$
to obtain $\Sigma_\sigma(\kk,\omega+i0^+) = \Sigma^\prime_\sigma + i \Sigma^{\prime\prime}_\sigma$.
This in turn leads to the spectral functions 
$A_\sigma(\kk,\omega) = - {\rm Im}
\left[\omega - \epsilon_\kk + \mu_\sigma - \Sigma^\prime_\sigma - i \Sigma^{\prime\prime}_\sigma\right]^{-1}/ \pi$,
which form the basis for our calculation of $n(\bf k)$ and of the RF~spectrum using (1). 

(I) Let us first discuss a simple self-consistent approximation
\cite{zwerger}, motivated by
an analysis that reproduces the essentially exact result \cite{combescot} 
of single $\dn$ spin ($n_{\dn} = 0$ limit) 
interacting with a Fermi sea of $\up$ fermions \cite{giorgini}. 
We will show that this scheme has  serious qualitative problems for $n_{\dn} > 0$ and analyze why this is the case. 
In this approximation, the Green's functions used to calculate $\Gamma$ and $\Sigma$
are the bare $G$'s but with a \emph{renormalized} chemical potential. A self-consistency 
condition is then imposed so that $\mu_\dn = \epsilon_{F\dn} + \Sigma^\prime_\dn(k_{F\dn},0;\mu_\dn)$,
where $\Sigma^\prime_\dn$ itself depends on $\mu_\dn$ \cite{combescot}. For the single minority spin limit, this
reproduces the result $\mu_\dn(n_{\dn} = 0) = - E_b \simeq - 0.6 \epsilon_{F\up}$.

\begin{figure}
\centerline{\epsfxsize=7.0truecm \epsfbox{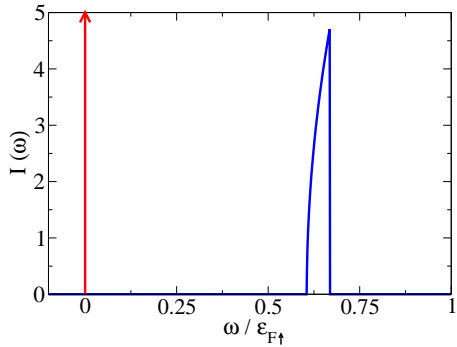}}
\caption{(color online) RF spectrum of a unitary Fermi gas
with $n_{\dn}/n_{\up} = 0.25$ calculated within the self consistent approximation (I) (see text). 
The majority ($\up$, red) is a delta function with weight $n_\up$. The  minority  ($\dn$, blue) spectrum has a discontinuity and a shift due to interactions, but no high frequency tail.}
\label{selfconsistent}
\end{figure}

This approximation for $n_{\dn} > 0$ implies the use of a \emph{negative} $\mu_\dn$
in the bare $G_\dn$ used in $\Gamma$ and $\Sigma$. As a result, one misses all effects of
finite $n_{\dn}$ occupancy ``inside'' the calculation. We can then analytically see that
${\rm Im}\Gamma(\qq,\omega < 0) \equiv 0$ which impacts the results
as follows. For the minority fermions $\Sigma^{\prime\prime}_\dn (\kk,\omega < 0) \equiv 0$,
which implies $A^{\rm inc}_\dn(\kk,\omega < 0) \equiv 0$ and thus $n_\dn(\kk) = Z_\dn\Theta(k_{F\dn} - k)$.
This means that $\sum_\kk n_\dn(\kk) = Z N_\dn < N_\dn$ and the zeroth moment sum rule for $I_\dn(\omega)$
is violated. In addition, in the absence of any incoherent spectral weight
for $\omega < 0$, one also misses both the universal $k^{-4}$ tail in $n_\dn(\kk)$
and the $\omega^{-3/2}$ tail in the RF spectrum (see Fig.~\ref{selfconsistent}).  
The first moment of $I_\dn(\omega)$ is then finite, instead of diverging as it should.
Further, the majority spins are \emph{completely unaffected by interactions} in this approximation
since one can see analytically that $\Sigma_\up (\kk,\omega) \equiv 0$, which is clearly
unphysical for non-zero $n_\dn$. The majority ($\uparrow$) RF spectrum is thus a delta function, a result that is 
at odds with all available experiments.  Clearly this approximation
fails to provide a reasonable description of RF spectra of highly imbalanced gases,
despite its success in obtaining reasonable numerical estimates for $\mu_\dn$.
All of the problems here arise from the fact that propagators
with renormalized $\mu_\dn < 0$ are used \emph{without} taking into the shifts in the $\dn$ particle dispersion. 

(II) This suggests that it may be physically more sensible to do the simplest
calculation \emph{without} any attempts at partial self-consistency, i.e., evaluate all
diagrams with bare propagators and bare $\epsilon_{F\sigma}$.
This leads to equations which are identical with the $1/N$ approximation \cite{sachdev}
with $N=1$ at the end. Now, in contrast to the previous approximation, the p-p vertex ${\rm Im}\Gamma$ has
structure even for $\omega < 0$, and this leads to $n(\kk)$ and $I(\omega)$ with
universal tails for both spins.

\begin{figure}
\centerline{\epsfxsize=7.0truecm \epsfbox{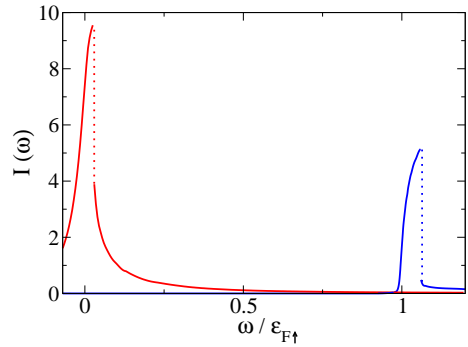}}
\caption{(color online) RF spectra of a unitary Fermi gas with $n_{\dn}/n_{\up} = 0.05$ within a simple ladder approximation (II) (see text). Both minority ($\dn$, blue) and majority ($\up$, red) spectra exhibit 
a discontinuity (dashed) and a large-$\omega$ tail. The blue curve for minority spins is  $15 \times I_{\downarrow}(\omega)$.   }
\label{rfspectrum}
\end{figure}

\begin{figure}
\centerline{\epsfxsize=7.0truecm \epsfbox{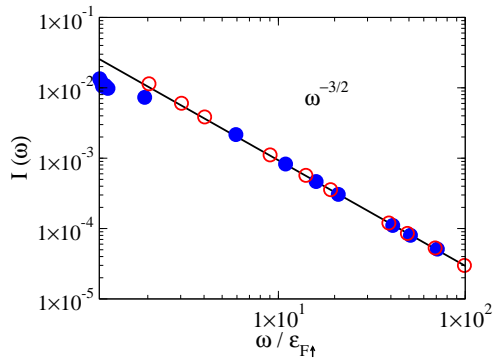}}
\caption{(color online) High frequency tails of the RF spectra of the unitary
Fermi gas with $n_{\downarrow}/n_{\uparrow} = 0.05$ shown in Fig.~\ref{rfspectrum}. 
Both majority (red, open circles) and minority (blue, filled circles) spectra exhibit a $\omega^{-3/2}$ tail. }
\label{tail}
\end{figure}

Our numerical results for the RF spectra of a highly polarized unitary Fermi gas
with $x = n_{\dn} /n_{\up} = 0.05$ are shown in Fig.~\ref{rfspectrum}.
Both majority and minority spectra show jump
discontinuities and high frequency tails. The
spin-independent $\omega^{-3/2}$ behavior tails are observed
in Fig.~\ref{tail}. Comparing the results of (I) the
self-consistent $\mu_\sigma$ approximation in Fig.~\ref{selfconsistent}
and (II) the simplest ladder approximation in 
Figs.~\ref{rfspectrum} and \ref{tail}, there is no doubt that the
latter provides a far better qualitative description of the RF spectrum.

Despite these qualitative successes, it must be emphasized that 
the simple ladder approximation (II) is not quantitatively accurate
insofar as the calculated chemical potentials, e.g., $\mu_\dn =
\epsilon_{F\dn} + \Sigma^\prime_\dn(k_{F\dn},0;\epsilon_{F\dn})$. In
particular we find that in the single spin limit $\mu_\dn(n_{\dn} = 0)
\simeq - 0.9 \epsilon_{F\up}$, as compared with the exactly result of
$- 0.6 \epsilon_{F\up}$. Moreover, we have found that the simple
ladder approximation leads to a negative compressibility for $x \gtrsim
0.05$ clearly signaling the limitations of the approximation.

The prospects for a better diagrammatic approximation are unclear, since
fully self-consistent calculations do not necessarily lead to better answers in
strongly interacting systems \cite{dsr}.
We also note that, while Quantum Monte Carlo (QMC) calculations have often provided valuable quantitative information\cite{giorgini} for energetics, 
extracting frequency-dependent correlation functions from QMC is very difficult
in view of two serious issues: the fermion sign problem in polarized systems 
and the problem of analytic continuation.

{\bf Comparison with Experiments:} Let us begin with the universal high frequency tail. A long tail is visible
in all of the published spectra (Fig.~1 of ref.~\cite{ketterle2008}; Fig.~2 of ref.~\cite{zwierlein2009}). It is
present for superfluid as well as normal state spectra and the same for both spin species, as we predict. 
It would be interesting to know if the signal to noise ratio in experiments is sufficient to test the $3/2$ power law \cite{final-state} and determine the coefficient $C$ \cite{normalization}. 

The jump discontinuity in the $T=0$ RF signal for a normal Fermi liquid will
be broadened by finite temperature and by experimental resolution. 
The best we can expect then is to see a peak at, or very close to, the location of the discontinuity. For the minority spins eq.~(\ref{disc}) predicts this to be 
$\omega^*_{\dn}/\epsilon_{F\up} = 3A_0/5 - (1 - m/m_0^*)x^{2/3} - 6Fx/5$,
using the best QMC result \cite{qmc} for $\mu_{\dn}$.
This is exactly the expression used for the peak position by Schirotzek {\it et al.} \cite{zwierlein2009}.
For the majority spins, with $m^* \simeq m$, the peak
will be at $\omega^*_{\up}/\epsilon_{F\up} = 2A_0x/5 + Fx^2/5$, which is slightly shifted from zero.

{\bf Conclusions:} We have derived two exact results for singular features in the RF spectra of
Fermi gases. The high frequency $\omega^{-3/2}$ is a universal feature, independent of the nature of the 
many-body state, when final state interactions are negligible, and provides an opportunity for measuring Tan's contact $C$. Such a study combined with other experimental probes such as photoassociation \cite{hulet2005} 
could provide deeper understanding of  how short range physics controls the universal properties of strongly interacting cold gases. Our second exact result on the jump discontinuity in the spectrum of a Fermi liquid at $T=0$ provides a distinguishing feature between a normal and superfluid ground state.
In the second part of our paper we show that, in the absence of a small parameter, it is very difficult to obtain reliable results for the detailed frequency dependence of the RF spectra
-- which capture both general qualitative features and are quantitatively accurate -- 
in strongly interacting quantum gases. Indeed, that makes exact results such as sum rules and
the singular features derived in this paper all the more important in interpreting experiments.

{\bf Acknowledgments}
We acknowledge discussions with R. Diener, W. Ketterle and M. Zwierlein,
and support from NSF and ARO. VBS thanks DST, India for  support through a Ramanujan grant.



\begin{thebibliography}{99}
\bibitem{normalization}
We will use the zeroth moment sum rule $N_\sigma$ to fix our normalization of $I_\sigma(\omega)$, 
since we have set the RF coupling constant in (1) to unity.
\bibitem{FLresults} The quasiparticle weight and effective mass are related to the real
self-energy $\Sigma^\prime$ via the standard results:
$Z = 1/\left[1 - \partial \Sigma^\prime/\partial \omega\right]$
and $m/m^* = Z \left[1 + \partial \Sigma^\prime/\partial \epsilon_k\right]$
evaluated at the Fermi surface: $(k=k_{F},\omega=0)$.
\bibitem{note}
If $m^* \simeq m$, but $Z < 1$, the singular contribution to $I(\omega)$ is a delta-function at 
$\omega^*$, and together with a incoherent piece which has the universal high frequency tail.
\bibitem{details} The $iq_\ell$ Matsubara sum can be done analytically using the spectral representation. 
The $\qq$-integrals in $\Sigma$ and $\kk$-integrals in (1) have to be done numerically.
\bibitem{qmc} We use 
$E = {3\over 5} N_\up \epsilon_{F\up}(1 - A_0x + (m/m_0^*)x^{5/3} + Fx^2)$,
the QMC parametrization \cite{giorgini} for the ground state energy. 
Here $x = n_{\dn} / n_{\up}$, $E_b = 3A_0 \epsilon_{F\up}/5$ is the binding energy and
$m_0^*$ the effective mass of a single $\dn$ spin, and
$F$ a Fermi-liquid correction.
We use $m_0^*$ here since $m^*$ is in general $x$-dependent.
\end{thebibliography}
\end{document}